\documentclass[11pt]{article}
\usepackage[font=small,labelfont=bf]{caption}

\usepackage{graphicx}
\usepackage{amssymb,rotating,enumerate,mathrsfs}
\usepackage[centertags]{amsmath}
\usepackage{color}
\usepackage[usenames,dvipsnames]{xcolor}

\usepackage{subfig}
\usepackage{hyperref}
\usepackage{mciteplus}

\def\avg#1{\langle #1 \rangle}

\def\a{\alpha}
\def\b{\beta}
\def\d{\delta}
\def\g{\gamma}
\def\G{\Gamma}
\def\e{\epsilon}
\def\et{\eta}
\def\L{\Lambda}
\def\m{\mu}
\def\n{\nu}
\def\O{\Omega}
\def\s{\sigma}
\def\r{\rho}

\def\t{\tau}

\def\ch{\chi}

\begin{document}

\begin{titlepage}

\begin{flushright}
UCB-PTH 12/08 \\
IPMU 12-0104 \\
\end{flushright}

\begin{center}

{\Large \bf Constraints on Light Dark Matter from Big Bang Nucleosynthesis}
\vskip 1cm

Brian Henning$^1$$^,$$^2$ and Hitoshi Murayama$^1$$^,$$^2$$^,$$^3$
\vskip 0.5cm

$^1$\emph{Department of Physics, University of California, Berkeley, California 94720, USA}

\vskip 0.3cm

$^2$\emph{Theoretical Physics Group, Lawrence Berkeley National Laboratory, Berkeley, California 94720, USA}

\vskip 0.3cm

$^3$\emph{Kavli Institute for the Physics and Mathematics of the Universe (WPI), Todai Institutes for Advanced Study, University of Tokyo, Kashiwa 277-8583, Japan}

\begin{abstract}
We examine the effects of relic dark matter annihilations on big bang nucleosynthesis (BBN). The magnitude of these effects scale simply with the dark matter mass and annihilation cross-section, which we derive. Estimates based on these scaling behaviors indicate that BBN severely constrains hadronic and radiative dark matter annihilation channels in the previously unconsidered dark matter mass region MeV \(\lesssim m_{\ch} \lesssim 10\) GeV. Interestingly, we find that BBN constraints on hadronic annihilation channels are competitive with similar bounds derived from the cosmic microwave background.
\end{abstract}

\end{center}
\end{titlepage}

\section{Introduction}
Dark matter populates our universe, representing more than 80\% of the total matter content. While astrophysical and cosmological measurements have elucidated the amount of dark matter (DM) in our universe as well as its role in galactic formation and distributions, we know very little about the nature of DM. For example, one of the few properties of DM that we can say anything with certainty is that dark matter has mass. But even the allowed mass values can span over 80 orders of magnitude, \(10^{-22} \text{ eV}\lesssim m_{\ch} \lesssim 10^{59} \text{ eV}\), with the lower bound coming from requirement of localization on galactic scales~\cite{Hu:2000ke} and the upper bound from micro-lensing searches for compact halo objects~\cite{Tisserand:2006zx}.

Despite our ignorance, we remain hopeful that theoretical models can provide a framework for experimentally probing dark matter directly and indirectly. For example, in this work we consider dark matter to be a weakly interacting massive particle that started its cosmic history in thermal equilibrium and froze-out as the universe cooled. The assumption of thermalization provides two powerful statements on DM: first, it restricts the allowed mass region to 11 orders of magnitude, \(\text{keV} \lesssim m_{\ch} \lesssim 100 \text{ TeV}\), with the lower and upper bounds coming from the requirement that dark matter be cold~\cite{Boyarsky:2008xj} and its annihilation unitary~\cite{Griest:1989wd}, respectively. Second, if DM is a thermal relic, its annihilations must freeze-out to reach the observed current abundance, providing an estimate on the strength of DM interactions. In particular, assuming annihilation is \(s\)-wave dominated, in order to meet the observed abundance, freeze-out requires a weak scale cross-section that is nearly independent of mass, \(\avg{\s v}_{\text{th}} \simeq 3 \times 10^{-26} \text{ cm}^3/\text{s}\).

After freeze-out, relic annihilations that occur may have observational consequences in two different scenarios. First, present day annihilations may be indirectly observed by searching the sky for their annihilation products (for a review, see~\cite{Cirelli:2012tf}). In the second scenario, relic annihilations can inject hadronic and/or electromagnetic energy that may alter events in our cosmic history, namely big bang nucleosynthesis (BBN) and recombination (CMB). Because the physics of nucleosynthesis and recombination are well understood, they offer particularly clean environments in which to probe non-standard effects, such as DM annihilation. In addition, changes to BBN and CMB by non-standard processes depend only on the type and rate of energy injection into the bath, and therefore leave BBN and CMB essentially decoupled from the high-energy details allowing relatively model-independent statements to be made.

In this work, we focus specifically on how relic annihilations from thermal dark matter affect nucleosynthesis. However, it is useful to describe in general how injected hadronic and electromagnetic energy alter nucleosynthesis and recombination. During nucleosynthesis, injection of hadronic and/or electromagnetic energy alters the abundances of nuclei via (i) energetic nucleons and photons dissociating nuclei and (ii) pions inter-converting background nucleons, thus raising the \(n/p\) ratio and therefore the primordial \({}^4\)He mass fraction. As for recombination, injected electromagnetic energy ionizes hydrogen atoms and therefore broadens the surface of last scattering. This results in scale dependent changes to the temperature and polarization power spectra, especially in the low \(l\) modes.

Now that we live in an era of precision CMB and BBN measurements, quantifying the effect of energy injection has led to powerful constraints on non-standard effects such as decay of long-lived particles~\cite{Reno:1987qw,Dimopoulos:1987fz,Cyburt:2002uv,Kawasaki:2004qu,Jedamzik:2004er,Jedamzik:2006xz,Chen:2003gz,Kasuya:2003sm,Zhang:2007zzh} and dark matter annihilation~\cite{Reno:1987qw,Frieman:1989fx,Jedamzik:2004er,Hisano:2009rc,Hisano:2008ti,Jedamzik:2009uy,Hisano:2011dc,Padmanabhan:2005es,Kanzaki:2009hf,Galli:2009zc,Slatyer:2009yq,Finkbeiner:2011dx,Natarajan:2012ry}. In the case of dark matter annihilation, the primary result obtained from analysis of the injected energy on nucleosynthesis and recombination is an upper bound on the DM annihilation cross-section, \(\avg{\s v}\), as a function of the DM mass. In particular, detailed analysis of the CMB has constrained \(s\)-wave annihilation to hadronic or electromagnetic channels to be less than the thermal cross-section required for freeze-out, \(\avg{\s v} < \avg{\s v}_{\text{th}}\), for \(m_{\ch} \lesssim 10\) GeV~\cite{Padmanabhan:2005es,Slatyer:2009yq,Finkbeiner:2011dx,Natarajan:2012ry}. Studies of \(s\)-wave annihilation on BBN have been performed by Hisano \emph{et. al.} for \(m_{\ch} \gtrsim 100\) GeV~\cite{Hisano:2009rc,Hisano:2008ti} and separately by Jedamzik and Pospelov for \(m_{\ch} \gtrsim \) several GeV~\cite{Jedamzik:2009uy}, while lower mass regions have remained unconsidered.

From previous works~\cite{Hisano:2009rc,Hisano:2008ti,Jedamzik:2009uy} that consider the effect of DM annihilation on BBN, the constraint on the annihilation cross-section appears to have a simple power law dependence on the DM mass, \(\avg{\s v} \propto m_{\ch}^{\d}\), where the power for hadronic (electromagnetic) energy injection is \(\d \approx 3/2\) (= 1). In this work, we explain these scaling behaviors and the range of DM masses for which they hold. Then we extrapolate the bounds to previously unconsidered DM masses and find that BBN constrains \(\avg{\s v} < \avg{\s v}_{\text{th}}\) for \(\text{few GeV} \lesssim m_{\ch} \lesssim 10 \text{ GeV}\) (\(30 \text{ MeV} \lesssim m_{\ch} \lesssim 75 \text{ MeV}\)) for the case of hadronic (electromagnetic) energy injection.

The structure of this paper is as follows: in section II we discuss how injection of hadronic and electromagnetic energy alters BBN, and in particular, how these effects scale with the dark matter mass and annihilation cross-section. In section III, we use these scaling laws along with results from precise numerical treatment~\cite{Hisano:2009rc} to estimate constraints on DM annihilation in a previously unconsidered low mass region. Our estimates indicate that BBN places a constraint on hadronic annihilation channels that is competitive and independent of the CMB constraint. We also discuss how these bounds change for annihilation to other Standard Model particles. Finally, in section IV we conclude and discuss future prospects for this work.

\section{Energy injection into BBN}
Injected energy from dark matter annihilations occurring after \(T \sim 1\) MeV can alter the primordial abundances of the light elements. Details of the evolution of injected energy and its effects on nucleosynthesis are thoroughly discussed in~\cite{Kawasaki:2004qu}. Quantifying the change on primordial abundances boils down to calculating the spectrum of produced hadron/nucleus \(H_i\). This spectrum, \(f_{H_i} = dn_{H_i}/dE_{H_i}\), is a complicated function that depends on the dynamics of the injected energy, the nuclear network leading to the production/destruction of \(H_i\), and the rate of DM annihilation. In this work, we focus on the dependence of this spectrum on the DM's mass and annihilation cross-section. Since production/destruction of \(H_i\) begins with an annihilation event, \(df_{H_i}/dt\) is proportional to the DM annihilation rate,
\begin{equation}
\frac{df_{H_i}}{dt} \propto \frac{dn_{\ch}}{dt} = - n_{\ch}^2 \avg{\s v}.
\label{eqn:spec_prop}
\end{equation}
Here, \(\avg{\s v}\) is the thermally-averaged annihilation cross-section which we take to be \(s\)-wave dominated and time independent and \(n_{\ch}\) is the number density of DM. For the range of DM masses considered, BBN occurs after freeze-out and therefore the number density of DM is fixed by observational abundance
\begin{equation}
n_{\ch}(t) = \frac{\O_{\ch}\r_{\text{crit}}}{m_{\ch}} \frac{1}{a(t)^3} \propto \frac{1}{m_{\ch}}.
\end{equation}
Hence, the DM injection rate scales as
\begin{equation}
\frac{d n_{\ch}}{dt} \propto - \frac{\avg{\s v}}{m_{\ch}^2}.
\label{eqn:ann_rate_prop}
\end{equation}
This has a simple physical interpretation: as dark matter mass decreases, in order to fit the observed abundance there needs to be more DM particles, which in turn leads to more annihilation events. In addition to the DM annihilation rate, the spectrum of produced/destroyed \(H_i\) will depend on the DM mass through the dynamics of the type of energy injected, which we now consider.

\subsection{Injection of hadronic energy}
Hadrons (\(p, n, \pi^{\pm},\) etc.) resultant from DM annihilation can primarily alter BBN through proton-neutron interconversion (\(\pi^{\pm}\)) and hadro-dissociation (\(p,n\))~\cite{Reno:1987qw,Kawasaki:2004qu}. Interconversion of \(p\) and \(n\) by energetic pions before the formation of D and \({}^4\)He (\(T \sim 100\) keV) can increase the \(n/p\) ratio set by \(\nu\) decoupling at \(T \sim \) MeV. Since almost all neutrons end up in \({}^4\)He (see, \emph{e.g.}~\cite{Kolb:1990vq}), the increase of \(n/p\) leads to a larger \({}^4\)He primordial mass fraction, \(Y_p\).

Energetic nucleons produced by DM annihilation alter primordial abundances through collisions with nuclei. The dominant effect of energetic neutrons and protons is to hadro-disassociate \({}^4\)He which in turn leads to an increased production of D, \({}^3\)H, \({}^3\)He, and \({}^6\)Li. Electromagnetic interactions with background photons and electrons tend to thermalize the nucleons, so hadro-dissociation only becomes efficient when the universe dilutes enough so that the electromagnetic interactions lose their stopping power. This occurs below \(T \sim 100\) keV for neutrons and \(T \sim 10\) keV for protons~\cite{Kawasaki:2004qu}. Because the primordial abundance of \({}^4\)He is much larger than other nuclides (for example, \(n_{\text{D}}/n_{{}^4\text{He}} \sim \mathcal{O}(10^{-3})\)), BBN is more sensitive to enhanced production of other nuclides than to processes which alter the primordial \({}^4\)He abundance. Therefore, hadro-dissociation of \({}^4\)He leading to increased production of other nuclides---namely D, \({}^3\)H, \({}^3\)He, and \({}^6\)Li---provides the most stringent constraints from BBN.


Since we wish to understand the change in a given nuclide's primordial abundance as a function of \(\avg{\s v}\) and \(m_{\ch}\), we consider how the amount of hadro-dissociation depends on these parameters. As in equation~\eqref{eqn:spec_prop}, the amount of hadro-dissociation is proportional to the dark matter annihilation rate, \(dn_{\ch}/dt \propto \avg{\s v}/m_{\ch}^2\). Since hadro-dissociation proceeds through energetic nucleons, further dependence on \(m_{\ch}\) essentially comes from the number of energetic nucleons produced in an annihilation event, \emph{i.e.} the nucleon multiplicity~\cite{Reno:1987qw,Dimopoulos:1987fz}.\footnote{Technically, it is the spectrum of nucleons per annihilation event, \(dN_N/dE_N\), that is of interest. The dependence of this spectrum on \(\sqrt{s} = 2m_{\ch}\) comes from the low \(x = E_N/\sqrt{s}\) behavior of the parton fragmentation functions \(D_i^N(x,\sqrt{s})\). The dependence of the multiplicity on \(\sqrt{s}\) shown in the text is inherited from \(dN_N/dE_N\)'s dependence on \(\sqrt{s}\)~\cite{Dokshitzer:1982xr,Biebel:2001ka}.} Assuming annihilation into quarks, \(\ch\ch\to q\bar{q}\), the nucleon multiplicity can be approximated by the multiplicity in \(e^+e^-\) collisions at \(\sqrt{s} = 2m_{\ch}\). To leading order in QCD, the average charged particle multiplicity has the following dependence on \(\sqrt{s}\)~\cite{Dokshitzer:1982xr,Biebel:2001ka,Webber:1984jp}:
\begin{equation}
\avg{n_{\text{ch}}} = a \exp\left( \sqrt{\frac{24}{\b_0}} \sqrt{\log\left( \frac{s}{\L^2} \right)} \right) + c
\end{equation}
where \(\b_0 = 11- 2n_f/3\), \(\L\) is the renormalization scale, and \(a\) and \(c\) are constants. Apart from overall normalization, for \(\sqrt{s} = 2m_{\ch}\) above several GeV, the above is well approximated by \((\sqrt{s})^{0.5 \pm \e} \sim m_{\ch}^{0.5 \pm \e}\) with \(\e \sim 0.05\).

The dependence of the amount of hadro-dissociation on the rate of DM annihilation and the nucleon multiplicity implies that the Boltzmann term for production/destruction of nuclei \(H_i\) approximately scales with DM as:
\begin{equation}
\left[\frac{dn_{H_i}}{dt}\right]_{\text{hadronic}} \propto m_{\ch}^{1/2} \frac{\avg{\s v}}{m_{\ch}^2} = \frac{\avg{\s v}}{m_{\ch}^{3/2}}
\label{eqn:hadronic_scaling}
\end{equation}
This scaling is expected to hold down to \(m_{\ch} \sim \) few GeV, below which hadro-dissociation quickly goes to zero since an annihilation event can no longer produce nucleons. This behavior implies that constraints on \(\avg{\s v}\) from BBN approximately scale as \(m_{\ch}^{3/2}\), which is confirmed by precise numerical calculations for \(m_{\ch} \gtrsim 10\) GeV~\cite{Hisano:2009rc,Hisano:2008ti,Jedamzik:2009uy} and shown in figure~\ref{fig:bounds}.

\subsection{Injection of electromagnetic energy}
Electromagnetic showering of energetic photons and leptons in the early universe produces many photons~\cite{Kawasaki:1994sc} which can photo-dissociate nuclei and alter primordial abundances~\cite{Cyburt:2002uv,Kawasaki:2004qu}. 

Photo-dissociation is a significant effect only when the energetic photons do not thermalize. Precise calculations of the evolution of electromagnetic cascades in the early universe can be found in the literature~\cite{Kawasaki:1994sc}, but the dominant effects are easily understood. Because of the small baryon-to-photon ratio, \(\et \sim \mathcal{O}(10^{-10})\), photons will pair produce off background photons (\(\g + \g_{\text{bkg}} \to e^+ + e^-\)) if their energy is above the threshold energy, \(E_{\text{th}} \approx m_e^2/22T\). Therefore, photo-dissociation only becomes efficient when the binding energy drops below the threshold for pair production.

Of the possible photo-dissociation channels, destruction of \({}^4\)He dominates since \({}^4\)He is the most abundant of the light nuclei by a few orders of magnitude. Since \(E_{\text{b,}{}^4\text{He}}\sim 20\) MeV, photo-dissociation of \({}^4\)He does not become relevant until \(T \lesssim 0.5\) keV. The dominant photo-dissociation channels of \({}^4\)He are to \({}^3\)H or \({}^3\)He plus a nucleon~\cite{Cyburt:2002uv}. Therefore, photo-dissociation of \({}^4\)He most noticeably leads to overproduction of \({}^3\)He (\({}^3\)H decays to \({}^3\)He with a half-life of about 14 years). Overproduction of D and \({}^6\)Li may also result from \({}^4\)He photo-destruction. In the case of deuterium, it can be directly produced in photo-spallation of \({}^4\)He~\cite{Cyburt:2002uv} or in capture of the neutron from \(\g + \a_{\text{bkg}} \to {}^3\text{He} + n\) on a background proton.
Non-thermal production of \({}^6\)Li may occur if the spallation product \({}^3\)H or \({}^3\)He is captured on a background \({}^4\)He, \({}^3\text{H} + \a_{\text{bkg}} \to {}^6\text{Li} + n\) or \({}^3\text{He} + \a_{\text{bkg}} \to {}^6\text{Li} + p\). Because the observed abundance of \({}^6\)Li is very small, non-thermal production of \({}^6\)Li via photo-dissociation of \({}^4\)He can have a noticeable impact~\cite{Jedamzik:1999di,Kawasaki:2000qr}.

Assuming annihilation into radiative channels (\(\ch\ch \to \g\g\) or \(\ch\ch\to l^+l^-\)), we wish to extract how production/destruction of nuclei \(H_i\) due to photo-dissociation depends on DM mass. As in the case of hadro-dissociation, the amount of photo-dissociation is proportional to the amount of DM annihilation, \(dn_{\ch}/dt\). The amount of photo-dissociation also depends on the number of photons produced in an annihilation event and the subsequent electromagnetic cascade. Because there is no mass gap for the photon, the number of photons produced is proportional to the visible energy (\(e^{\pm}\) and \(\g\)) from the annihilation event, \(E_{\text{vis}}\)~\cite{privNom:2012}. For example, annihilation to electrons or muons have \(E_{\text{vis}} = 2m_{\ch}\) and \(E_{\text{vis}} \sim 2m_{\ch}/3\), respectively. 

\begin{figure}[t]
\centering
\includegraphics[width=0.95\textwidth]{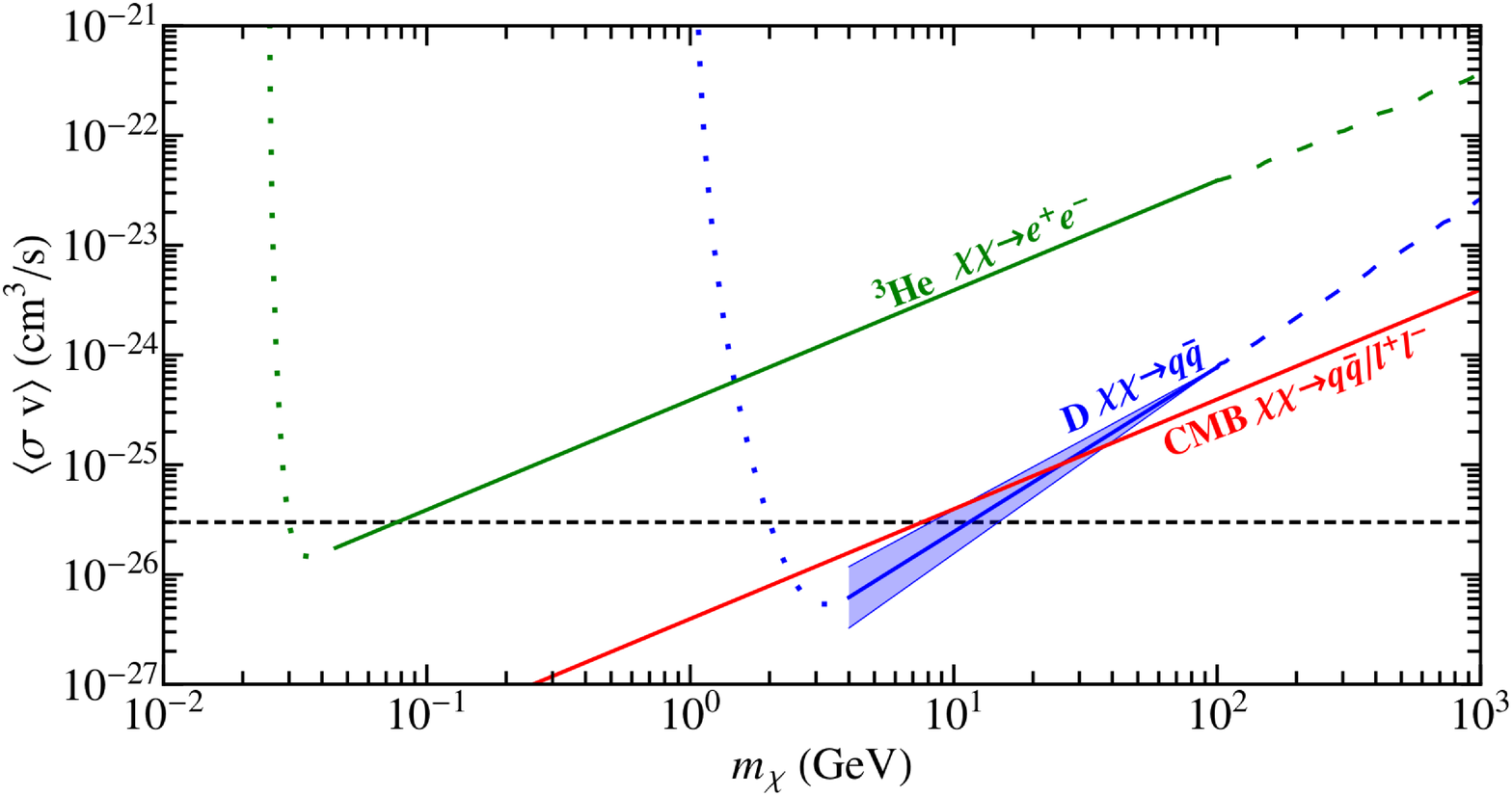}
\caption{BBN constraints on hadronic (blue) and radiative (green) annihilation channels from D and \({}^3\)He overproduction, respectively. The solid blue (green) line is our estimate based on the scaling behavior \(\avg{\s v} \propto m_{\ch}^{1.5 \pm 0.2}\) (\(\avg{\s v}\propto m_{\ch}\)) for hadronic (radiative) annihilation. The long dashed lines indicate bounds on \(\ch\ch\to b\bar{b}\) (blue) and \(\ch\ch \to e^+e^-\) (green) from Hisano \emph{et. al.}~\cite{Hisano:2009rc}. The dotted blue and green lines is a rendering of where we expect these scaling laws to break down, but their exact shape and placement requires precise numerical treatment. The red line is the bound from the CMB, whose normalization is set by constraints on \(\ch\ch\to b\bar{b}\) at 95\% C.L. from Natarajan~\cite{Natarajan:2012ry} and extended using the scaling \(\avg{\s v} \propto m_{\ch}\)~\cite{Padmanabhan:2005es,Kanzaki:2009hf,Galli:2009zc,Slatyer:2009yq,Finkbeiner:2011dx,Natarajan:2012ry}. For CMB, the important quantity is visible energy injection, and therefore annihilation to quarks or charged leptons give roughly the same constraint.}
\label{fig:bounds}
\end{figure}

The amount of photo-dissociation implies that the Boltzmann term for production/destruction of nuclei \(H_i\) due to photo-dissociation is inversely proportional to the DM mass,
\begin{equation}
\left[\frac{dn_{H_i}}{dt}\right]_{\text{P.D.}} \propto \frac{\avg{\s v}}{m_{\ch}} \left(\frac{E_{\text{vis}}}{m_{\ch}}\right).
\label{eqn:EM_scaling}
\end{equation}
Therefore, if DM annihilates into radiative channels, constraints on \(\avg{\s v}\) from BBN scale as \(m_{\ch}\), which is confirmed by precise numerical calculations for \(m_{\ch} \gtrsim 10\) GeV~\cite{Hisano:2009rc,Jedamzik:2009uy,Hisano:2011dc} and shown in figure~\ref{fig:bounds}. This scaling can be expected to hold down close to the binding energy of \({}^4\)He, \(E_{b,{}^4\text{He}} \sim 20\) MeV. In the case of non-thermal production of \({}^6\)Li this scaling holds down to \(\sim 60\) MeV. This is because the daughter \({}^3\)H or \({}^3\)He from photo-dissociation of \({}^4\)He needs about 10 MeV kinetic energy to efficiently capture on background \({}^4\)He~\cite{Cyburt:2002uv}, thus requiring \(E_{\g}\gtrsim 60\) MeV. Note that constraints from the CMB also exhibit the scaling behavior \(\avg{\s v} \propto m_{\ch}\)~\cite{Padmanabhan:2005es,Kanzaki:2009hf,Galli:2009zc,Slatyer:2009yq,Finkbeiner:2011dx,Natarajan:2012ry} since the amount of visible energy from the annihilation is what dictates how many photons will be produced.

\section{Constraints from BBN}
Relic dark matter annihilations alter the primordial abundances of nuclei from standard BBN predictions. In the case of \(s\)-wave annihilation, as discussed in the previous section, the scaling of these effects on DM mass and annihilation cross-section is given in equations~\eqref{eqn:hadronic_scaling} and~\eqref{eqn:EM_scaling} for hadronic and radiative annihilation channels, respectively. 

In this work, we use BBN to place an upper bound on \(\avg{\s v}\) for the previously unconsidered low DM mass region of \(\text{MeV} \lesssim m_{\ch} \lesssim 10 \text{ GeV}\). To do this we adopt the bound on \(\avg{\s v}\) at \(m_{\ch} = 100 \text{ GeV}\) from Hisano \emph{et. al.}~\cite{Hisano:2009rc} and extrapolate the bounds to the low mass region using~\eqref{eqn:hadronic_scaling} and~\eqref{eqn:EM_scaling}.

While a precise treatment requires numerical simulation, results from other works~\cite{Hisano:2009rc,Hisano:2008ti,Jedamzik:2009uy,Hisano:2011dc} indicate that the scaling behavior in~\eqref{eqn:hadronic_scaling} and~\eqref{eqn:EM_scaling} is robust. We will discuss where this scaling behavior breaks down and how precise our estimates are.

Since we adopt the constraints of Hisano \emph{et. al.}~\cite{Hisano:2009rc} as a starting point, the observational abundances used to place bounds are the same as in that work. These observational abundances are discussed in appendix~\ref{app:obs_abund}.

\subsection{Hadronic constraints}
As discussed in equation~\eqref{eqn:hadronic_scaling}, constraints on \(\avg{\s v}\) from hadronic energy injection are approximately proportional to \(m_{\ch}^{2-\a}\) where \(\a \approx 0.5\) comes from the dependence of the nucleon multiplicity on \(\sqrt{s} = 2m_{\ch}\); \(\avg{n_N}\) is approximately proportional to \(m_{\ch}^{\a}\). As \(m_{\ch}\) approaches a GeV, this scaling is expected to break down and constraints on \(\avg{\s v}\) to rapidly weaken since annihilations no longer produce energetic nucleons that dissociate background \({}^4\)He.

The scaling \(\avg{\s v} \propto m_{\ch}^{1.5}\) assumes the dominant effect on primordial abundances is hadro-dissociation of background \({}^4\)He. While hadro-dissociation does dominate, there may be other processes which contribute a small but non-negligible amount. For example, DM annihilations may further enhance primordial D, \({}^3\)He, and \({}^6\)Li abundances through (1) photo-dissociation of background \({}^4\)He by energetic photons produced in the hadronic jet and from electromagnetic energy loss and (2) more background \({}^4\)He to dissociate from \(p-n\) interconversion by pions at \(T \gtrsim 100\) keV. While these contributions are sub-dominant to hadro-dissociation of background \({}^4\)He, precise numerical treatment is needed to evaluate the relative size and scaling with \(m_{\ch}\) of all possible effects. The scaling behavior with \(m_{\ch}\) might also be modified by the number of secondary energetic nucleons produced from hadro-dissociation processes, although this effect is subdominate~\cite{Kawasaki:2004qu}.

Our goal in this work is to derive conservative estimates. Therefore, to parametrize our ignorance on the interplay of all effects in calculating constraints on \(\avg{\s v}\) for a given mass, we include an uncertainty in the scaling
\begin{equation}
\avg{\s v} \propto m_{\ch}^{1.5 \pm \d}
\label{eqn:had_scaling_uncert}
\end{equation}
where we take the uncertainty to be \(\d = 0.2\)~\cite{Hisano:2009rc}.

In Ref~\cite{Hisano:2009rc} the authors consider annihilation into \(b\) quarks, \(\ch\ch\to b\bar{b}\). From the primordial deuterium abundance they derive an upper bound \(\avg{\s v} \le 7.7\times 10^{-25} \text{ cm}^3/\text{s}\) for \(m_{\ch} = 100\) GeV. As shown in figure~\ref{fig:bounds}, the scaling~\eqref{eqn:had_scaling_uncert} implies \(\avg{\s v} < \avg{\s v}_{\text{th}} = 3\times 10^{-26} \text{ cm}^3/\text{s}\) for
\begin{equation}
m_b \lesssim m_{\ch} \le 11.4_{-3.2}^{+3.3} \text{ GeV}.
\label{eqn:had_constraint}
\end{equation}
For annihilation to bottom quarks, the constraint is cutoff at \(m_b\sim 5\) GeV. While the above constraint is for \(\ch\ch\to b\bar{b}\), similar results are anticipated for annihilation to other quark species. However, in the case of annihilation to lighter quarks, we expect the constraint on \(\avg{\s v}\) falls off as \(m_{\ch} \to m_N \sim \text{ GeV}\), but where and how rapidly this occurs can only be obtained from numerical analysis. 

\subsection{Electromagnetic constraints}
Constraints on \(\avg{\s v}\) from electromagnetic energy injection are proportional to \(m_{\ch}\). Electromagnetic energy primarily alters primordial abundance through photo-dissociation of \({}^4\)He, leading to enhanced production of other nuclides. Therefore, the scaling \(\avg{\s v}\propto m_{\ch}\) holds as long as the DM annihilation and subsequent shower produce photons energetic enough to dissociate \({}^4\)He.

As discussed earlier, the observable consequences of overproduction as a result of \({}^4\)He photo-dissociation are largest for \({}^3\)He. In~\cite{Hisano:2009rc}, Hisano \emph{et. al.} study the annihilation channel \(\ch\ch\to e^+e^-\) and derive an upper bound of \(\avg{\s v} \le 3.9 \times 10^{-23} \text{ cm}^3/\text{s}\) for \(m_{\ch} = 100\) GeV from \({}^3\)He overproduction. As shown in figure~\ref{fig:bounds}, the scaling \(\avg{\s v} \propto m_{\ch}\) implies \(\avg{\s v} < \avg{\s v}_{\text{th}} = 3 \times 10^{-26} \text{ cm}^3/\text{s}\) for
\begin{equation}
30 \text{ MeV} \lesssim m_{\ch} \le 77 \text{ MeV}
\label{eqn:EM_constraint}
\end{equation}

The constraints fall off when the electrons from the DM annihilation can no longer produce photons energetic enough to dissociate \({}^4\)He. The photo-dissociation cross-section for \({}^4\)He, \(\s_{{}^4\text{He}\g\to\dots}\), proceeds through the giant dipole resonance and only becomes efficient when \(E_{\g} \gtrsim 25\) MeV~\cite{Cyburt:2002uv,Calarco:1983zza}. An electron from the DM annihilation event with \(E_{e^{\pm}} = m_{\ch}\) near this threshold produces energetic photons via inverse-Compton scattering off background photons. Inverse-Compton scattering is a relatively hard event, with \(\sim 80\%\) of the momentum transferred to the scattered photon. Therefore we expect the constraints on \(\ch\ch\to e^+e^-\) from \({}^3\)He overproduction to fall off around 30 MeV.

\subsection{Other annihilation channels}
The constraints~\eqref{eqn:had_constraint} and~\eqref{eqn:EM_constraint} for hadronic and electromagnetic energy injection are obtained from DM annihilation to \(b\bar{b}\) and \(e^+e^-\), respectively. It is useful to consider how these constraints would change for different annihilation channels. Nucleosynthesis is sensitive to the amount of hadronic and electromagnetic energy injected into the bath and is therefore essentially decoupled from the high energy details of the dark matter annihilation. Therefore, in order to understand how different DM annihilation channels alter BBN, it is sufficient to specify the annihilation products and ask what particles are produced in the resulting hadronic and/or electromagnetic cascade.

In the case of annihilation into hadronic channels, we do not anticipate the constraints from different quark species (\(u,d,s,c\)) to change much from the case of annihilation to bottom quarks. Injected hadronic energy primarily alters BBN through dissociation of nuclei by energetic nucleons resultant from the hadronization of the initial state quarks. At energies much larger than the nucleon mass, the spectrum of nucleons does not significantly depend on the initial state quark. As the DM mass gets near the nucleon mass, \(E\gtrsim m_N\), the nucleon spectrum from hadronization will differ between initial state quark species. For lighter quark species, the main change from the constraint for \(\ch\ch\to b\bar{b}\)~\eqref{eqn:had_constraint} is that the bounds will fall off at a DM mass lower than \(m_b\sim 5\) GeV. Therefore, for annihilation to quarks, we generically expect \(\avg{\s v}_{\ch\ch\to q\bar{q}} < \avg{\s v}_{\text{th}}\) for \(\text{few GeV} \lesssim m_{\ch} \lesssim 11 \pm \text{few} \text{ GeV}\) where the lower limit comes from the inability to produce energetic nucleons in the annihilation event and will have a slight dependence on initial quark species, beyond the precision of these estimates.

Annihilation to photons, electrons, or muons will alter BBN through the injection of electromagnetic energy. From equation~\eqref{eqn:EM_scaling}, constraints on the DM annihilation cross-section are inversely proportional to the amount of injected visible energy (electrons and photons),
\begin{equation}
\avg{\s v} \propto m_{\ch} \left( \frac{m_{\ch}}{E_{\text{vis}}} \right).
\end{equation}
In the case of annihilation to photons or electrons \(E_{\text{vis}} = 2m_{\ch}\). For annihilation to muons, \(E_{\text{vis}} \sim 2m_{\ch}/3\) and therefore the constraints on \(\avg{\s v}\) for \(\ch\ch\to \m^+\m^-\) are weakened by a factor of \(\sim 3\) from the case of \(\ch\ch \to e^+e^-\). Therefore, we estimate that \({}^3\)He overproduction bounds \(\avg{\s v}_{\ch\ch\to \m^+ \m^-} \lesssim 1 \times 10^{-25} (m_{\ch}/m_{\m}) \text{ cm}^3/\text{s}\) and is valid for \(m_{\ch} \ge m_{\m}\).

BBN constraints on annihilation to taus, \(\ch\ch \to \t^+\t^-\), come from the injection of electromagnetic energy from the \(\t\) decay products. Despite the large hadronic branching fraction in the \(\t\) decay, since \(m_{\t} < 2 m_{N}\), annihilation to \(\t\)'s cannot produce nucleons which would alter nucleosynthesis through hadro-dissociation of \({}^4\)He. While the hadronic products of \(\t\) decay (pions and kaons) will slightly affect the primordial \({}^4\)He mass fraction through interconversion of background \(n\) and \(p\) at \(100 \text{ keV} \lesssim T \lesssim \text{MeV}\), they predominantly alter standard BBN through the electromagnetic energy in their decay products. Taking the average fraction of visible energy in a \(\t\) decay to be 0.31~\cite{Hisano:2009rc}, overproduction of \({}^3\)He from photo-dissociation of \({}^4\)He bounds \(\avg{\s v}_{\ch\ch \to \t^+\t^-} \lesssim 1 \times 10^{-24} (m_{\ch}/10 \text{ GeV}) \text{ cm}^3/\text{s}\).

Finally, we comment on annihilation to neutrinos. Injection of energetic neutrinos during nucleosynthesis has been considered in the context of a long-lived particle decay~\cite{Gratsias:1990tr,Kawasaki:1994bs,Kanzaki:2007pd}. While neutrinos have impact nucleosynthesis much less than colored or charged particles, here we estimate the magnitude of effects on nucleosynthesis from annihilation to neutrinos and show that the effects are, at most, as large as \(\mathcal{O}(10^{-4})\) the effect of radiative annihilation channels. High energy neutrinos may pair produce charged leptons off of background neutrinos, \(\n + \bar{\n}_{\text{bkg}} \to l^+ + l^-\). If this reaction happens after \(T \sim 0.5\) keV, the photons in the electromagnetic shower of the charged leptons can photo-dissociate nuclei and alter their primordial abundances~\cite{Gratsias:1990tr,Kawasaki:1994bs,Kanzaki:2007pd}. Charged leptons are also generically produced in the annihilation through radiation of a (real or virtual) weak boson by one of the final state neutrinos, \emph{e.g.} \(\ch\ch\to \n_i \bar{l}_i W^{(*)}\to \n_i \bar{l}_i \bar{\n}_jl_j \).\footnote{Of course, if the primary annihilation channel is through neutrinos, annihilation to \(l^+l^-\) is generated at loop order and is model-dependent. Given a model, this annihilation can be calculated and the bounds for \(\ch\ch\to l^+l^-\) used. We focus on radiation of a weak boson giving a three- or four-body final state since it is model-independent.} 

Both processes---high energy neutrinos pair producing off the neutrino background and annihilation to a multi-body final state---are important for charged lepton production, and therefore for BBN constraints. For charged lepton pair production off a background neutrino the rate of interaction is approximately \(\G_{\n \bar{\n}_{\text{bkg}} \to l^+l^-} \sim G_F^2 m_{\ch} T_{\n} n_{\n,\text{bkg}}\). This reaction can occur as long as the initial neutrino is above the threshold for pair production, \(E_{\n}E_{\n_{\text{bkg}}} = m_{\ch}T_{\n} \ge m_{l}^2\), which for production of \(e^+e^-\) occurring after \(T \sim 0.5\) keV requires \(m_{\ch} \gtrsim \) GeV. The fraction of neutrinos that pair produce after \(T \sim 0.5\) keV is then approximately\footnote{Cosmological redshift for the energetic neutrino can actually occur since \(\G_{\n \bar{\n}_{\text{bkg}} \to l^+l^-} < H\). However, to not complicate our point, we do not include it in the estimate. Cosmological redshift lowers the fraction of neutrinos that pair produce charged leptons, but does not change our conclusions.}
\begin{equation}
\int_{t(T\sim0.5\text{ keV})}^{t_f} dt \ \G_{\n \bar{\n}_{\text{bkg}} \to l^+l^-} \sim 2\times10^{-4}\left( \frac{m_{\ch}}{100 \text{ GeV}} \right) \left( \frac{T}{0.5 \text{ keV}} \right)^2.
\end{equation}
Note that if the initial neutrino is energetic enough, high energy secondary neutrinos from neutrino-neutrino scattering can play a role~\cite{Kawasaki:1994bs,Kanzaki:2007pd}. As for the multi-body final state, the relative amount of charged lepton production via neutrino radiation of a weak boson is approximately \(G_F^2 m_{\ch}^4/16\pi^2\) for \(m_{\ch} \ll M_W\) and can be as large as \(\a_w/4\pi \sim 3\times10^{-3}\) when the weak boson is radiated on shell. Thus, depending on the energy of the initial neutrino, both the high energy neutrinos themselves and branching to multi-body final states can have the same order-of-magnitude effect on nucleosynthesis, which is what was found in~\cite{Kanzaki:2007pd}. Based on these considerations, we estimate that bounds on annihilation to neutrinos are weaker than bounds on radiative annihilation channels by \(\mathcal{O}(10^{-4}-10^{-6})\) for \(m_{\ch} \gtrsim \)GeV and by \(\sim G_F^2 m_{\ch}^4/16\pi^2\) for \(m_{\ch} \lesssim \) GeV. For this reason, annihilation to neutrinos is essentially negligible for BBN.

\section{Conclusions and Future Work}
In this work we have examined the effects of dark matter annihilations during the epoch of big bang nucleosynthesis. We emphasize that, in terms of the annihilation itself, the magnitude of these effects depend only on the rate of energy injection (\emph{i.e.} the DM annihilation rate) and the type of energy injected (hadrons, charged leptons, \emph{etc.}). With this procedure, we have explained how changes to primordial abundances of nuclei scale with the dark matter mass and annihilation cross-section. These scaling behaviors are robust and have been observed in precise numerical treatments of DM annihilation during BBN~\cite{Hisano:2009rc,Hisano:2008ti,Jedamzik:2009uy,Hisano:2011dc}.

The dependence of changes to nucleosynthesis on the dark matter mass and annihilation cross-section, along with results from precise numerical calculation~\cite{Hisano:2009rc}, have allowed us to estimate constraints on \(\avg{\s v}\) for the previously unconsidered low DM mass region \(\text{MeV}\lesssim m_{\ch} \lesssim 10 \text{ GeV}\). Interestingly, our estimates indicate that BBN rules out generic \(s\)-wave annihilation to quarks (radiative \(e^+e^-\),\(\g\g\)) for \(\text{few GeV}\lesssim m_{\ch} \lesssim 10 \text{ GeV}\) (\(30 \text{ MeV}\lesssim m_{\ch} \lesssim 500 \text{ MeV}\)).

Our results have focused on the case that the thermally averaged annihilation cross-section is independent of time (\(s\)-wave annihilation). For scenarios in which \(\avg{\s v}\) depends on time, changes to the time-independent case can be understood from a modification of the rate of annihilation, \(\G_{\text{ann}} = n_{\ch}(t)\avg{\s v}(t)\). For example, if the annihilation is \(s\)-wave suppressed, \(\avg{\s v}\) will decrease in time~\cite{Kolb:1990vq} and therefore have less impact on BBN. On the other side, as studied in~\cite{Hisano:2011dc}, Sommerfeld~\cite{Hisano:2003ec} or Breit-Wigner~\cite{Ibe:2008ye} type enhancements lead to \(\avg{\s v}\) increasing in time and having a stronger impact on nucleosynthesis.

Besides making our estimates precise, our results warrant a full numerical treatment, which is currently in progress~\cite{WIPHenn:2012}, for a few reasons. Our estimates indicate that \(s\)-wave annihilation to quarks is ruled out starting around \(m_{\ch} \sim 10\) GeV, which is competitive with the current bound from CMB~\cite{Finkbeiner:2011dx,Natarajan:2012ry}. In estimating constraints for light DM, we extrapolated the results of Hisano \emph{et. al.}~\cite{Hisano:2009rc}, whose philosophy was to provide conservative constraints. Given that BBN provides an independent, and possibly more stringent, constraint than CMB, it is worthwhile to perform a full numerical calculation that also quantifies the confidence limit on the constraints placed.

\section*{Acknowledgments}
We would like to thank M.~Kawasaki, K.~Kohri, and T.~Moroi for helpful comments and discussions. B.H. is grateful to W.~Haxton, E.~Mereghetti, Y.~Nomura, and M.~Papucci for useful conversations. This work was supported in part by the U.S. DOE under Contract DE-AC03-76SF00098, and in part by the NSF under grant
PHY-1002399.  The work by H.M. was also supported in part by the JSPS grant (C) 23540289, in part by the FIRST program “Subaru Measurements of Images and Redshifts (SuMIRe)”, CSTP, Japan, and by WPI, MEXT, Japan.

\appendix

\section{Observational abundances of light elements}
\label{app:obs_abund}
Since we use the results of Ref~\cite{Hisano:2009rc} as a starting point for our analysis, we adopt the observed primordial abundances used in their analysis.

The primordial D/H abundance is inferred from the QSO absorption line in metal poor systems,
\begin{equation}
(n_{\text{D}}/n_{\text{H}})_{\text{p}} = (2.82 \pm 0.26) \times 10^{-5}.
\end{equation}
This value, used in Ref~\cite{Hisano:2009rc}, is the weighted average of six observed QSO absorption systems. A more recent work by the same group includes a seventh measurement that doesn't change the central value, but lowers the dispersion to \(\pm 0.20 \times 10^{-5}\)~\cite{Hisano:2011dc}.

An upper bound on the primordial \({}^3\)He abundance is obtained from \({}^3\)He/D measurements in protosolar clouds,
\begin{equation}
(n_{{}^3\text{He}}/n_{\text{D}})_{\text{p}} < 0.83 + 0.27.
\end{equation}

While we do not use Li to place bounds in this work, we comment here on its measurement and the so-called lithium problem. \({}^7\)Li is observed in the atmospheres of metal-poor Population II stars in our galactic halo. Recently, \({}^6\)Li has been observed in the these systems as well~\cite{Asplund:2005yt}, although the total number of systems with observable \({}^6\)Li is somewhat controversial due to observational difficulties in distinguishing \({}^6\)Li and \({}^7\)Li spectra~\cite{Cayrel:2008hk}.

Both the \({}^7\)Li and \({}^6\)Li measurements conflict with the theoretical values predicted by standard BBN and are collectively referred to as the lithium problem (for a recent review, see~\cite{Fields:2012jf}). For \({}^7\)Li, the observed abundance is lower by a factor of about three than the theoretical value predicted by standard BBN, with a \(4-5\s\) significance. In the case of \({}^6\)Li, the observed abundance is more than a factor of \(10^2\) larger than the standard BBN prediction. While the solution to the lithium problems may be unknown, a conservative approach, as used in~\cite{Hisano:2009rc}, is to take the observed \({}^6\)Li and \({}^7\)Li abundances as upper limits for their primordial abundances. There is also an extra uncertainty in the primordial abundances that comes from the possibility that stellar burning may deplete primordial \({}^6\)Li and \({}^7\)Li~\cite{Pinsonneault:1998nf}. In this case, the conservative approach is to include this depletion as an uncertainty that raises the primordial abundance~\cite{Hisano:2009rc}.


\end{document}